\documentclass{article}

\RequirePackage{hyperref}
\usepackage[utf8]{inputenc} %unicode support

\usepackage{ams math, amssymb, booktabs,graphicx}
\usepackage{rotating}

\newcommand{\Prob} {\mbox{$\rm{Prob}$\,}}

\title{Estimation of the methylation pattern distribution from deep sequencing data}
\author{Peijie Lin$^1$, Sylvain For\^et$^2$, Susan R.\ Wilson$^{1, 3}$\\ and Conrad J.\ Burden$^1$}
\date{}

\begin{document}
\maketitle

\vspace{-0.5cm}
\begin{center}
\begin{small}
$^1$Mathematical Sciences Institute, Australian National University, Canberra \\
$^2$ Research School of Biology, Australian National University, Canberra\\
$^3$ School of Mathematics and Statistics, University of New South Wales, Sydney\\
\vspace{0.5cm}
\end{small}
\end{center}

\begin{abstract} % abstract

Motivation: Bisulphite sequencing enables the detection of cytosine methylation.
The sequence of the methylation states of cytosines on any given read forms a
methylation pattern that carries substantially more information than merely
studying the average methylation level at individual positions.
In order to understand better the complexity of DNA methylation landscapes in
biological samples, it is important to study the diversity of these methylation
patterns.
However, the accurate quantification of methylation patterns is subject to
sequencing errors and spurious signals due to incomplete bisulphite conversion
of cytosines.  

Results: A statistical model is developed which accounts for the distribution of DNA
methylation patterns at any given locus.
The model incorporates the effects of sequencing errors and spurious reads, and enables 
estimation of the true underlying distribution of methylation patterns.  

Conclusions: Calculation of the estimated distribution over methylation patterns is
implemented in the R Bioconductor package {\tt MPFE}. 
Source code and documentation of the package are also available for download at
http://bioconductor.org/packages/3.0/bioc/html/MPFE.html. 

\end{abstract}

%%%%%%%%%%%%%%%%%%%%%%%%%%%%%%%%%%%%%%%%%%%%%%
%%                                          %%
%% The keywords begin here                  %%
%%                                          %%
%% Put each keyword in separate \kwd{}.     %%
%%                                          %%
%%%%%%%%%%%%%%%%%%%%%%%%%%%%%%%%%%%%%%%%%%%%%%

%%%%%%%%%%%%%%%%%%%%%%%%%%%%%%%%%%%%%%%%%%%%%%
%%                                          %%
%% The Main Body begins here                %%
%%                                          %%
%% Please refer to the instructions for     %%
%% authors on:                              %%
%% http://www.biomedcentral.com/info/authors%%
%% and include the section headings         %%
%% accordingly for your article type.       %%
%%                                          %%
%% See the Results and Discussion section   %%
%% for details on how to create sub-sections%%
%%                                          %%
%% use \cite{...} to cite references        %%
%%  \cite{koon} and                         %%
%%  \cite{oreg,khar,zvai,xjon,schn,pond}    %%
%%  \nocite{smith,marg,hunn,advi,koha,mouse}%%
%%                                          %%
%%%%%%%%%%%%%%%%%%%%%%%%%%%%%%%%%%%%%%%%%%%%%%

%%%%%%%%%%%%%%%%%%%%%%%%% start of article main body
% <put your article body there>

%%%%%%%%%%%%%%%%
%% Background %%
%%
\section*{Introduction}

Epigenetic regulations are involved in a broad range of biological processes,
including development, tissue homeostasis, learning and memory, as well as
various diseases such as obesity and cancer
\cite{Cantone2013, Day2010, Feinberg2007}.

DNA methylation is one of the best studied epigenetic molecular mechanisms.
It consists of the addition of a methyl group to the cytosine residues (C) of a
DNA molecule.
In animals, DNA methylation usually takes place in the CpG context: cytosines
followed by a guanine (G) residue.

DNA methylation modulates gene expression through a variety of mechanisms.
In vertebrates, methylation in the promoter region usually has a repressive
effect on transcription initiation.
By contrast, methylation of gene bodies is generally associated with an active
transcriptional state and has been shown to play an important role in the
control of alternative splicing \cite{LFK+10, Shukla2011}.

The diverse and subtle effects of DNA methylation enable a given genome to
produce different phenotypic outputs as part of a developmental program or in
response to environmental factors.
This has fundamental implications at the organismal level, where DNA
methylation plays an important role in phenotypic plasticity
\cite{Kucharski2008}.
This is also important at the cellular level to create diverse cell types,
tissues and organs all based on the same genome.
DNA methylation patterns can thus change from one cell type to another or
within a cell under different conditions \cite{Hawkins2010}.

The diversity of methylation patterns in a sample can be studied with a single
base pair resolution using the bisulphite sequencing
technique~\cite{Clark1994}.
When DNA is treated with bisulphite, the unmethylated cytosines are converted
to uracils with a high (albeit not complete) efficiency, whereas the methylated
cytosines remain as cytosines.
A library is prepared from the bisulphite treated DNA by fragmenting to lengths
of approximately 200~bp and PCR amplified.
During this amplification process, uracils are replicated as thymines (T).
The DNA library is then sequenced and the resulting reads are mapped to a
reference.
Within each read, CpG dinucleotides which have been converted to TpG are
recognised as unmethylated, and unconverted CpG dinucleotides are recognised as
methylated.

A common type of analysis carried out with this type of data is to estimate,
for each cytosine, the global methylation level of a sample, namely, the
average methylation level across all DNA strands in all the cells represented
in that sample.
This is usually done by correcting for the fact that bisulphite conversion is
not a complete reaction (for instance, see \cite{LFK+10,Akman14}).

However, this type of analysis only considers the methylation level at
individual positions and is oblivious to the fact that cytosines present on a
given read represent a broader snapshot of the methylation landscape on a
particular strand of DNA.
One can therefore gain a significantly deeper insight into the complexity of
DNA methylation landscapes by reconstituting the methylation patterns that are
physically present on the same sequence, and thus come from the same cell.
Each read can be seen as containing a small number of binary labelled CpG sites:
1 for methylated, 0 for unmethylated and represents the methylation pattern of
a given strand of DNA in one particular cell.
This approach is of particular interest when studying complex biological
samples that contain a mixture of cell types, for instance a tumour or a whole
insect brain.
It gives a much more detailed picture about the diversity of DNA methylation in a
sample than simply looking at the methylation level at each position.

The approach of looking at methylation patterns instead of individual CpG sites 
can be hampered by the lack of sequencing coverage depth.
For reasons of cost, most whole genome bisulphite sequencing studies have a
mean genomic coverage in the 10-100 range.
If the underlying sample contains hundreds of methylation patterns, they cannot
be sampled representatively.

As an alternative, in order to reach a high coverage, researchers have often
focussed on specific loci, either by sequencing PCR amplicons of bisulphite
converted DNA, or by reduced representation bisulphite sequencing, or by using a
capture assay (for review see \cite{Lee:2013aa}).
In this paper, we focus on the analysis of amplicon data, but our method is
widely applicable to other sequencing approaches.

The problem of determining methylation patterns in the original sample from the
observed data is non trivial.
The purpose of this paper is to infer formally the probability distribution
over all possible methylation profiles defined by the population of epigenomes
at a given locus.

\section*{Results}

\subsection*{Synthetic data}
\label{sec:Synthetic}

The probability distribution over methylation patterns at a given locus from 
from bisulphite sequencing data is estimated via an algorithm described below 
in the Methods section. The algorithm is
implemented as an R Bioconductor~\cite{Bioconductor04} package {\tt MPFE}
(for \textbf{M}ethylation \textbf{P}atterns \textbf{F}requency \textbf{E}stimation).

As the true distribution over methylation patterns is always unknown in the laboratory, we have
constructed a number of synthetic data sets to test the effectiveness of the
algorithm. 
For a locus with $n$ CpG sites we prescribe a true distribution $\theta_k$, 
$k = 1, \ldots, 2^n$, over the $2^n$ possible methylation patterns.
We simulate from these patterns a set of $N_{\rm read}$ initial reads, each
labelled with its `true' methylation pattern, and then redistribute the reads
among patterns according to the statistical model described in 
the Methods section to simulate errors due to incomplete bisulphite
conversion and sequencing errors.  
Incomplete conversion is specified by a non-conversion rate parameter
$\epsilon$ equal to the probability that an unmethylated cytosine will fail to
convert.
Its value is typically of order $10^{-2}$, and will be assumed in our
simulations to be estimated independently from the rate of conversion of
non-CpG cytosines.  
For instance, it could be calculated based on sequences known a priori not to
be methylated.
The sequencing error rates are specified by a vector
$\eta = (\eta_1, ..., \eta_n)$,
where $\eta_s \sim O(10^{-2})$ is the probability that site $s$ is methylated
but registers as unmethylated and vice versa. 
These two sources of error can generate spurious methylation patterns, that is,
patterns for which the observed read count $y_k$ is non-zero when the true
frequency of the pattern, $\theta_k$, is zero.
The aim is to assess the ability of the algorithm to recover the true
distribution $\theta_k$ from a set of observed read counts $y_k$, and, more
specifically, determine whether spurious patterns can be identified. 

The algorithm can be run in two possible modes: a slow mode, which performs a
complete calculation over all possible $2^n$ methylation patterns, and a
(default) fast mode which assumes $\theta_k = 0$ for any pattern for which the
observed read count $y_k$ is zero.
Because of the exponential growth in the number of patterns, we find in general
that use of the function in the slow mode becomes computationally prohibitive for $n > 8$.
However, there is very little difference in the computed results between the
two modes (see below).
Furthermore, the number of realised patterns tends to be relatively small and
the performance of the function is adequate even for large $n$.  

\begin{figure}
\centering
\includegraphics[width=\textwidth]{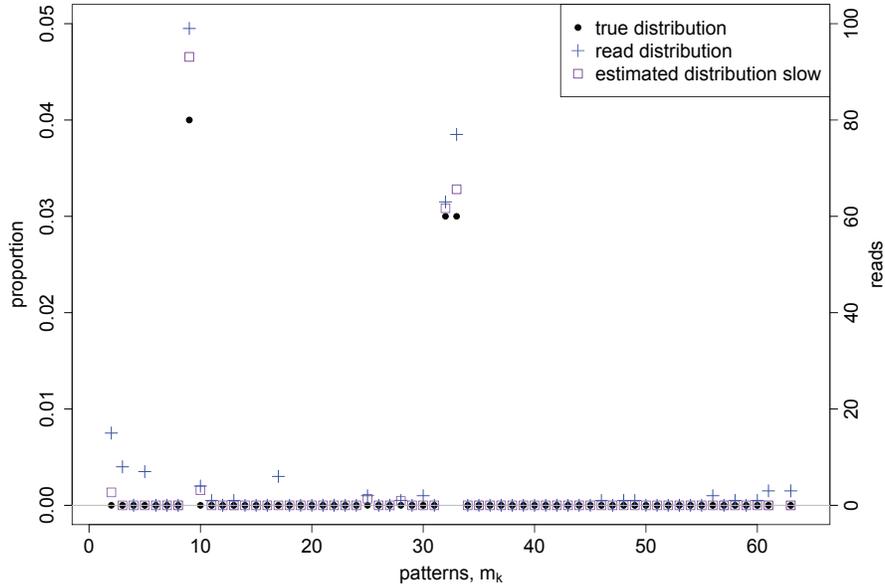}
\caption{The `true' distribution $\theta_k$ (black dots), observed read counts $y_k$ (blue pluses) and 
 estimated distribution $\hat\theta_k$ (purple squares) over $2^n$ methylation 
 patterns for synthetic data with $n = 6$ CpG sites.  The non-conversion rate is set to be $\epsilon = 0.005$, 
 the sequencing errors are set to be $\eta=(0.008, 0.006, 0.006, 0.006, 0.006, 0.008)$,
 the total number of reads is $N_{\rm read} = 2000$ and the estimated distribution was calculated using the 
 slow version of the algorithm which estimates a distribution over all $64$ possible patterns.  
 Methylation patterns are labelled lexicographically 
 from $m_1 = 000000$ to $m_{64} = 111111$.  
 Data for patterns $m_1$, $m_{62} = 111101$, and $m_{64}$ are beyond the range of the plot,  
 but are listed in Table~\ref{syntheticTable}.  
}
\label{sdslowf}
\end{figure}

Figure~\ref{sdslowf} shows the results of analysing synthetic data in the slow
mode at a locus with $n = 6$ CpG sites, a non-conversion rate $\epsilon
=0.005$, sequencing errors $\eta=(0.008, 0.006, 0.006, 0.006, 0.006, 0.008)$ to
model a signal that degrades towards the ends of the reads, and a
total number of reads $N_{\rm read} =2000$.  

\begin{table}
\centering
\caption{Comparison between estimates $\hat\theta_i$ of the true methylation distribution $\theta_i$ calculated with the slow and fast
implementations of our algorithm for the dataset of Figure \ref{sdslowf}.  $y_i$ is the number of observed reads for pattern $i$ 
and $N_{\rm read}$ is the total number of reads.\label{syntheticTable}}
\vspace{5mm}
{\begin{tabular}{l  r r c l l}\hline
Patterns  &	$\theta_i$ & $y_i$ & $y_i/N_{\rm read}$ &  $\hat\theta_i$ (slow) & $\hat\theta_i$ (fast)	\\\hline
000000    &  0.50  & 907  &  0.4535     &  0.4813    &  0.4812	\\
000001    &  0.00  &  15  &  0.0075     &  0.0013   &  0.0013	\\
000010    &  0.00  &   8  &  0.0040     & 0    		& 0		\\
000100    &  0.00  &   7  &  0.0035     & 0    		& 0		\\
001000    &  0.04  &  99  &  0.0495     &  0.0466    &  0.0466	\\
001001    &  0.00  &   4  &  0.0020     &  0.0016    &  0.0014	\\
001010    &  0.00  &   1  &  0.0005     & 0    		& 0		\\
001100    &  0.00  &   1  &  0.0005     & 0    		& 0		\\
010000    &  0.00  &   6  &  0.0030     & 0   			 & 0		\\
011000    &  0.00  &   2  &  0.0010     &  0.0007    &  0.0004	\\
011011    &  0.00  &   1  &  0.0005     &  0.0005    &  0.0002	\\
011101    &  0.00  &   2  &  0.0010     & 0    		& 0		\\
011111    &  0.03  &  63  &  0.0315     &  0.0308    &  0.0306	\\
100000    &  0.03  &  77  &  0.0385     &  0.0328    &  0.0329	\\
101101    &  0.00  &   1  &  0.0005     & 0   			& 0		\\
101111    &  0.00  &   1  &  0.0005     & 0    		& 0		\\
110000    &  0.00  &   1  &  0.0005     & 0    		& 0.0000	\\
110111    &  0.00  &   2  &  0.0010     & 0    		& 0		\\
111001    &  0.00  &   1  &  0.0005     & 0   		 	& 0		\\
111011    &  0.00  &   1  &  0.0005     & 0    		& 0		\\
111100    &  0.00  &   3  &  0.0015     & 0    		& 0		\\
111101    &  0.20  & 393  &  0.1965     &  0.2001    &  0.2013	\\
111110    &  0.00  &   3  &  0.0015     & 0    		& 0		\\
111111    &  0.20  & 401  &  0.2005     &  0.2044    &  0.2040	\\
\hline
\end{tabular}}{}
\end{table}

The results for the patterns with a non-zero number of observed reads are also
listed in Table~\ref{syntheticTable}.  
We observe that for almost all patterns the estimated distribution
$\hat\theta_k$ is closer to the true distribution $\theta_k$ than the naive
read proportion $y_k/N_{\rm read}$, the only exception being a slight shift in
the wrong direction for the pattern $m_{64} (111111)$.  
The algorithm correctly identifies $14$ out of the $18$ spurious patterns.
Note that our implementation of the algorithm registers whether, for any given
pattern $k$, the estimate $\hat\theta_k$ is identically zero, that is, it makes
a call as to whether the pattern is present or absent.
In this example, no real pattern is classified as spurious.

\begin{figure}
\includegraphics[width=\textwidth]{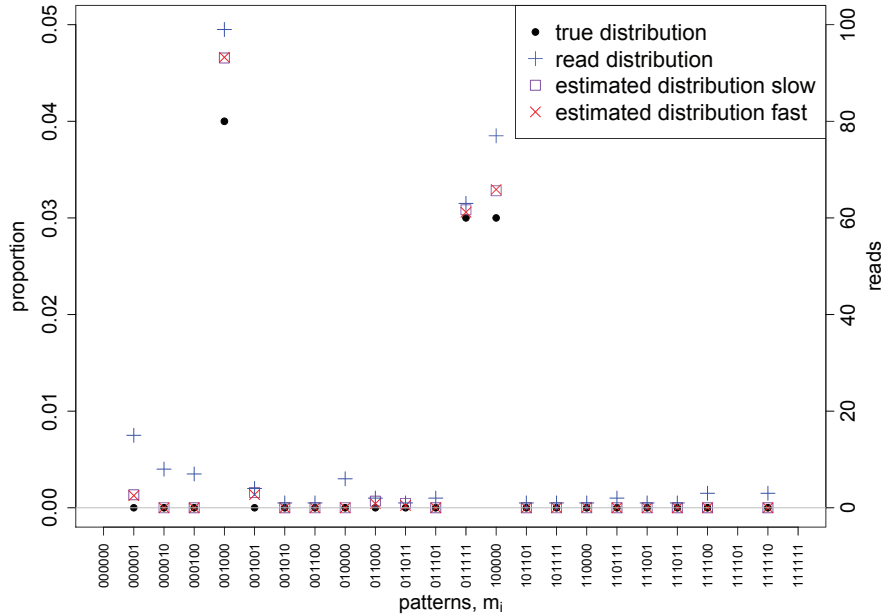}
\caption{Comparison between estimates $\hat\theta_i$ calculated with the slow (purple squares) and fast (red crosses) 
implementations of our algorithm for the dataset of Figure \ref{sdslowf}.  
Only patterns with non-zero reads are shown.
Data for the three patterns ($000000$, $111101$ and $111111$) which are beyond the range of the plot can be found in Table~\ref{syntheticTable}.}
\label{swco}
\end{figure}

Table \ref{syntheticTable} and Figure \ref{swco} illustrate the comparison
between the slow and (default) fast implementations of the algorithm applied to
the data which were analysed using the slow version in Figure \ref{sdslowf}.  
In general, when the number of CpG sites $n$ in an amplicon is large, it is
expected that only a small faction of the $2^n$ possible patterns will be
present.
Furthermore, it is rare for a true pattern with $\theta_k > 0$ to
have zero counts as a result of incomplete bisulphite conversion or sequencing
errors.
For instance, in the example of Figure \ref{sdslowf} none of the patterns with
a positive true frequency $\theta_k$  had zero reads.
For the remaining patterns, i.e. those plotted in Figure \ref{swco}, there is
no substantial difference between the two implementations of the algorithm.
The fast implementation identifies one less spurious pattern than the slow implementation; 
that is pattern $110000$, however, it estimates a very low proportion, with $\hat{\theta}_i \mbox{(fast)} < 10^{-4}$. 
From now on, the discussion focusses on the fast implementation.

\begin{figure}
\includegraphics[width=\textwidth]{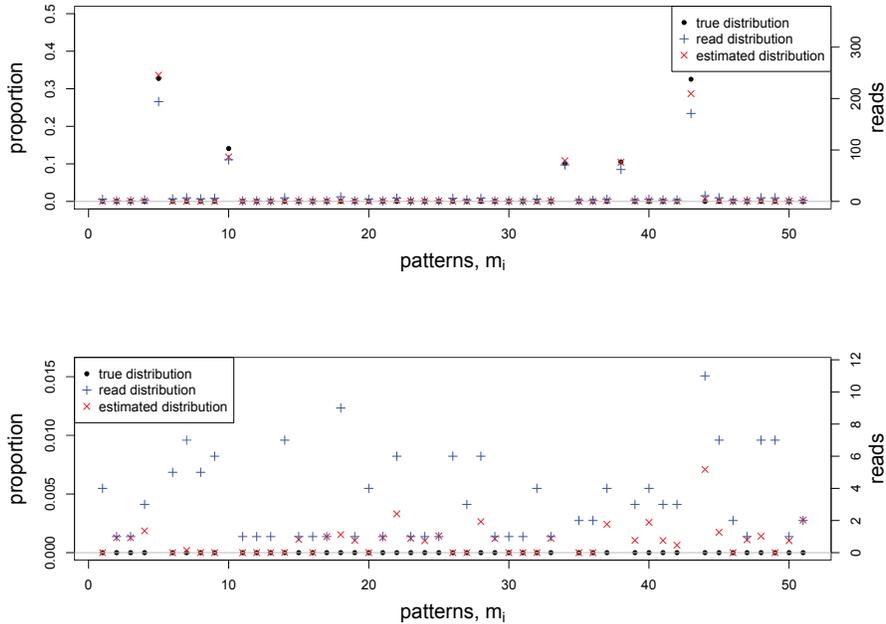}
\caption{Comparison between the true distribution $\theta_i$, the simulated read distribution $y_i/N_{\rm read}$, and the estimated 
distribution $\hat\theta_i$ for a synthetic dataset based on amplicon data.
Parameter values are $n = 9$ CpG sites, total number of reads $N_{\rm read} = 730$, 
non-conversion parameter $\epsilon = 0.01$, and sequencing error rate $\eta = 0.02$.  
The lower plot is an expanded version of the upper plot showing only patterns with low frequencies.}
\label{syntheticBee}
\end{figure}

Figure~\ref{syntheticBee} shows the results of applying the fast implementation
of the algorithm to synthetic data modelled on biological data from a PCR amplicon
in the honey bee {\it Apis mellifera} genome (see Biological data section).
To obtain the dataset, the function was applied once to a biological dataset of
$N_{\rm read} = 730$ reads from an amplicon corresponding to a locus with $n =
9$ CpG sites.  
To maintain a similar number of non-zero reads the `true' distribution of the synthetic dataset was taken to be 
$$
\theta_i = 
\begin{cases} 
0 & \text{if } \hat\theta_i^{\rm init} < 0.05, \\
\hat\theta_i^{\rm init} & \text{otherwise}, 
\end{cases}
$$
where $\hat\theta_i^{\rm init}$ is the result of the initial application of the algorithm.    
Here the non-conversion parameter is $\epsilon = 0.01$, and the sequencing
error rate $\eta = 0.02$ is taken to be uniform across all CpG sites.  

Of the $46$ spurious patterns in this dataset, $18$ are correctly
identified and no false identifications of spurious patterns are made. 
For the $28$ failed identifications, the program generally estimates a lower
estimate $\hat\theta_i$ than the read proportion $y_i/N_{\rm read}$.

\subsection*{Hypothesis testing}

The effectiveness of our algorithm in identifying spurious patterns can be further gauged in terms of classical hypothesis testing.  
In the following definitions we only consider methylation patterns with non-zero reads $y_i$.

We set a threshold $K \in [0, 1]$ and, using the estimated distribution
$\hat\theta_i$ as a test statistic, declare pattern $i$ to be spurious when
$\hat\theta_i \leq K$.
Patterns are defined to be true or false positives or negatives according to
the rules

$$
\mbox{TP if } \hat\theta_i \leq K \mbox{ and } \theta_i = 0, \qquad
\mbox{FP if } \hat\theta_i \leq K \mbox{ and } \theta_i > 0,
$$
$$
\mbox{TN if } \hat\theta_i > K \mbox{ and } \theta_i > 0, \qquad
\mbox{FN if } \hat\theta_i > K \mbox{ and } \theta_i = 0.
$$
True positive rates (TPR) and false positive rates (FPR) are defined in the usual way as 
$$
\mbox{TPR} = \frac{\mbox{TP}}{\mbox{TP} + \mbox{FN}}, \qquad
\mbox{FPR} = \frac{\mbox{FP}}{\mbox{FP} + \mbox{TN}}.  
$$
An analogous set of definitions applies using the raw data count proportions
$y_i/N_{\rm read}$ as a test statistic.  

\begin{figure}
\includegraphics[width=\textwidth]{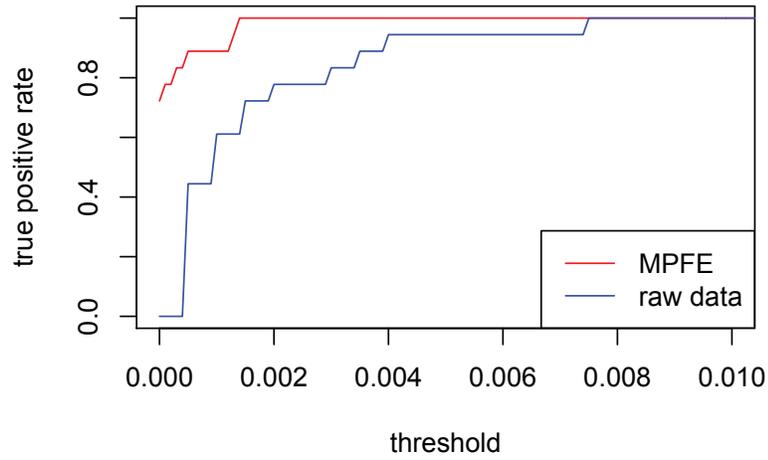}
\caption{The TPR versus the threshold $K$ 
using the test statistic $\hat\theta_i$ (red curve) and $y_i/N_{\rm read}$ (blue curve) for the data of Figure \ref{swco}.}
\label{tpr}
\end{figure}

Figure~\ref{tpr} shows the TPR curves for the data of
Figure~\ref{swco}.  
It shows that using $\hat{\theta}_i$ results in a clear improvement in detecting which methylation patterns
are likely to be a spurious artefact of incomplete conversion and reading
error. 
The FPR curves for both test statistics are
constantly zero for the same threshold range in the TPR graph.

\begin{figure}
\includegraphics[width=\textwidth]{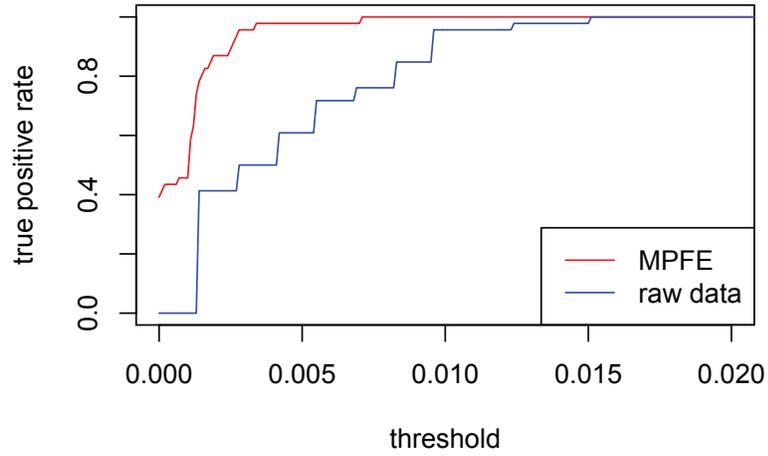}
\caption{The TPR versus the threshold $K$ using the test statistic
$\hat\theta_i$ (red curve) and $y_i/N_{\rm read}$ (blue curve) for the
synthetic data analysed in Figure~\ref{syntheticBee}.}
\label{real_tpr}
\end{figure}

Figure~\ref{real_tpr} shows the TPR curves for the synthetic data based on
biological data from an amplicon analysed in Figure~\ref{syntheticBee}.  
Again we observe a clear improvement in detecting which methylation patterns
are likely to be a spurious artefact of incomplete methylation and reading
error.  

\subsection*{Biological data}

Amplicons were obtained as described in \cite{LFK+10}.
Briefly, genomic DNA was extracted from brains of adult honeybee workers and treated with sodium bisulphite.
A region of gene GB17113 (gene ID: 724724, a 6-phosphofructokinase) was then amplified by PCR and sequenced using the 454 technology.

\begin{figure}
\includegraphics[width=\textwidth]{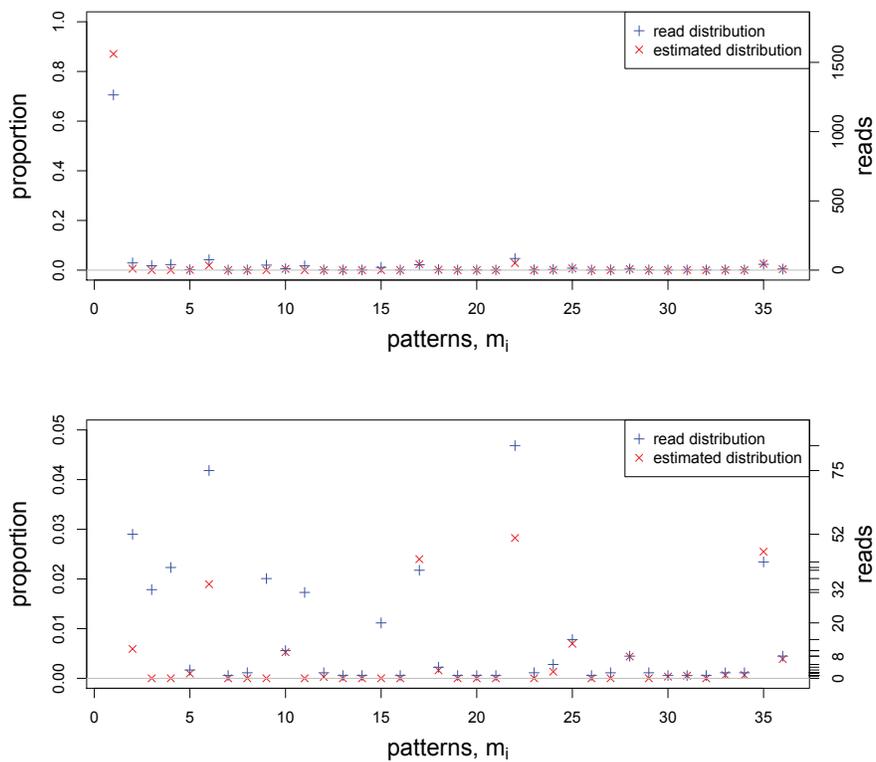}
\caption{Estimated distribution $\hat{\theta}_i$ of the methylation patterns 
obtained for a honey bee amplicon.
Parameter values are $n=8$ CpG sites, 
total number of reads $N_{\rm read}=1793$, 
non-conversion rate $\epsilon = 0.01$, 
and sequencing error rate $\eta=0.02$.
The lower plot is an expanded version of the upper plot 
showing only patterns with low frequencies.
The 36 patterns with non-zero reads are listed in Table~\ref{biologicalTable}.}
\label{biologicalData2}
\end{figure}

Below we apply the fast algorithm to two examples from this dataset assuming a
non-conversion rate of $\epsilon = 0.01$ and a global sequencing error rate of
$\eta = 0.02$.  
The first example is shown in Figure~\ref{biologicalData2} and
Table~\ref{biologicalTable}.  
The parameter values are $n=8$ CpG sites, $36$ patterns with non-zero reads,
and a total number of reads $N_{\rm read}=1793$.

\begin{table}
\centering
\caption{Estimated methylation distribution $\hat\theta_i$ for the 
data of Figure~\ref{biologicalData2}.
The 36 patterns with non-zero reads are 
labelled $m_i$, $i = 1, \ldots, 36$ and $y_i$ are the observed read counts.\label{biologicalTable}}
\vspace{5mm}
{\begin{tabular}{r c  r c l}\hline
$i$   &  $m_i$ &	$y_i$  &  $y_i/N_{\rm read}$ 	 &	$\hat\theta_i$	\\\hline
1   &  00000000 &	 1265  &  0.7055       &  0.8706			\\
2   &  00000001 &	   52  &  0.0290       &  0.0059			\\
3   &  00000010 &	   32  &  0.0178       & 0				\\
4   &  00000100 &	   40  &  0.0223       & 0				\\
5   &  00000110 &	    3  &  0.0017       &  0.0010			\\
6   &  00001000 &	   75  &  0.0418       &  0.0189			\\
7   &  00001001 &	    1  &  0.0006       & 0				\\
8   &  00001010 &	    2  &  0.0011       & 0				\\
9   &  00010000 &	   36  &  0.0201       & 0				\\
10  &  00010100 &	   10  &  0.0056       &  0.0053			\\
11  &  00100000 &	   31  &  0.0173       & 0				\\
12  &  00100001 &	    2  &  0.0011       &  0.0003			\\
13  &  00101000 &	    1  &  0.0006       & 0				\\
14  &  00110000 &	    1  &  0.0006       & 0				\\
15  &  01000000 &	   20  &  0.0112       & 0				\\
16  &  01000001 &	    1  &  0.0006       & 0				\\
17  &  01100000 &	   39  &  0.0218       &  0.0240			\\
18  &  01100001 &	    4  &  0.0022       &  0.0016			\\
19  &  01101000 &	    1  &  0.0006       & 0				\\
20  &  01110000 &	    1  &  0.0006       & 0				\\
21  &  01110100 &	    1  &  0.0006       & 0				\\
22  &  10000000 &	   84  &  0.0468       &  0.0282			\\
23  &  10000001 &	    2  &  0.0011       & 0				\\
24  &  10000010 &	    5  &  0.0028       &  0.0013			\\
25  &  10000100 &	   14  &  0.0078       &  0.0070			\\
26  &  10001000 &	    1  &  0.0006       & 0				\\
27  &  10010000 &	    2  &  0.0011       & 0				\\
28  &  10010100 &	    8  &  0.0045       &  0.0044			\\
29 	&  10100000 &	    2  &  0.0011       & 0				\\
30  &  11000100 &	    1  &  0.0006       &  0.0003			\\
31  &  11010000 &	    1  &  0.0006       &  0.0005			\\
32  &  11100000 &	    1  &  0.0006       & 0				\\
33  &  11100100 &	    2  &  0.0011       &  0.0006			\\
34  &  11110000 &	    2  &  0.0011       &  0.0007			\\
35  &  11110100 &	   42  &  0.0234       &  0.0255			\\
36  &  11110110 &	    8  &  0.0045       &  0.0039			\\
\hline
\end{tabular}}{}
\end{table}

There are several observations:
\begin{enumerate}
\item[\rm (i)] $18$ patterns ($50\%$) are identified as spurious;
\item[\rm (ii)] there are $11$ patterns with only $1$ read - our algorithm calls $9$ of them as spurious, while predicts the other $2$ patterns ($m_{30}$ and $m_{31}$) to exist;
\item[\rm (iii)] patterns $m_3$, $m_4$, $m_9$ and $m_{11}$ with $> 30$ reads each has a read proportion $y_i/N_{\rm read} \approx 2\%$, but are called as spurious. 
\end{enumerate}

Observations (i) and (ii) can be explained by the fact that the edit distance between the two patterns covered by a single read and any pattern observed to be highly 
abundant renders it unlikely that these patterns have arisen through sequencing errors or incomplete conversion. 
Observation (iii) arises because the spurious patterns with $> 30$ 
reads are just one sequencing error or one incomplete conversion away from the most abundant pattern, $m_0$(00000000).

\begin{sidewaysfigure}
\includegraphics[width=\textwidth]{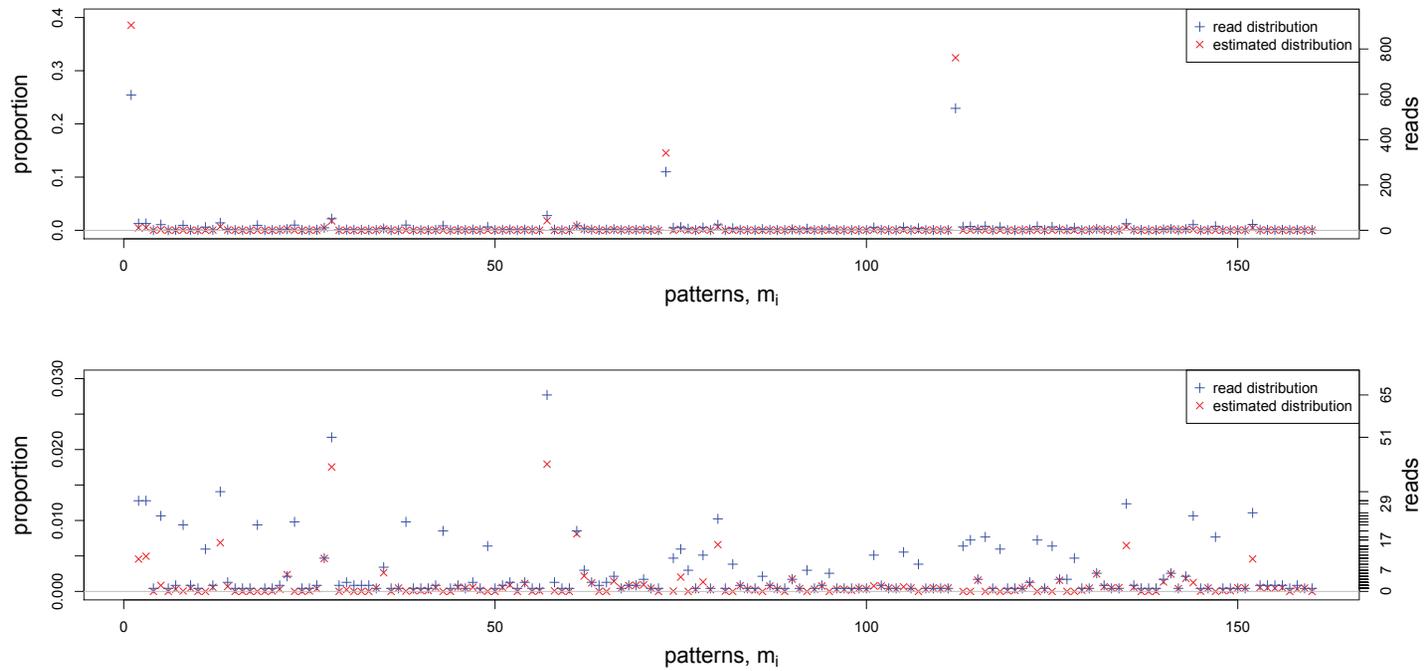}
\caption{Estimated distribution $\hat{\theta}_i$ of the methylation patterns
for a second amplicon from the honey bee genome.
Parameter values are $n=14$ CpG sites, 
total number of reads $N_{\rm read}=2347$, 
non-conversion rate $\epsilon = 0.01$, 
and sequencing error rate $\eta=0.02$.
The lower plot is an expanded version of the upper plot showing only patterns
with low frequencies.}
\label{biologicalData1}
\end{sidewaysfigure}

Figure \ref{biologicalData1} shows the second example, with $n=14$ CpG sites and $160$ patterns with
non-zero reads. 
The total number of reads is $N_{\rm read}=2347$.
In this case $47$ patterns are called as spurious.

\section*{Methods}
\label{sec:methods}

\subsection*{Statistical model of bisulphite sequencing}
\label{sec:statModel}

We take as the starting point of our statistical model a population of
epigenomes restricted to a locus containing $n$ CpG sites.
Each member of the population within a given class is represented by a vector
of non-independent binary valued random variables $\mathbf{K} = (K_1, \ldots,
K_n)$, where each $K_s \in \{0, 1\}$ labels the methylation state (1 for
methylated, 0 for unmethylated) at the $s$-th CpG site at this locus.
The population defines a methylation profile represented by the probability
distribution of realising the pattern $\mathbf{k} = (k_1, \ldots, k_n)$ in a
read randomly chosen from the population:
\begin{equation}
\theta_\mathbf{k} = \Prob(\mathbf{K} = \mathbf{k}), \quad \mathbf{k} \in \{0, 1\}^n.   \label{probK}
\end{equation}
For convenience, from here on we will label the possible methylation patterns
by the integers $k = 1, \ldots, 2^n$, and set $\theta_\mathbf{k} = \theta_k$,
where $k - 1$ is the integer whose binary representation is the methylation
pattern $\mathbf{k}$.  

Our aim is to estimate the distribution $\theta_k$ representing the relative
abundance of methylation pattern $k$ from high throughput sequencing data
consisting of a set of integer valued read counts $Y_k$.
In a typical experiment, the number $n$ of CpG sites in an amplicon may be up
$O(10^2)$, and the total number of read counts $N_{\rm read} = \sum_{k =
1}^{2^n} Y_k$ may be up to $O(10^6)$.  

The model takes into account two sources of error.
First, the bisulphite conversion of unmethylated cytosine to uracil is not
100\% efficient.
There is a probability $\epsilon$ that an unmethylated CpG site will register
as being methylated, where $\epsilon \sim O(10^{-2})$ can be estimated from the
cytosines known not to be methylated (mitochondrial genome, chloroplastic
genome, spike-ins, etc).
The second type of error is caused by sequencing.
For many applications this may be assumed for practical purposes to be site
independent.
However, to allow for effects such as degradation of the read quality towards
the ends of the reads, we will assume there is a site-dependent probability
$\eta_s \sim O(10^{-2})$ that if site $s$ is unmethylated it will register as
methylated and vice versa.  
It follows that if the true methylation pattern of any given read is
$\mathbf{K}$, but the read registers as being pattern $\mathbf{L}$, then 
\begin{equation}
\Prob(L = \ell | K = k) = M_{k \ell}, \quad k, \ell = 1, \ldots 2^n,      \label{probLgivenK}
\end{equation}
where
\begin{equation}
M = \bigotimes_{s = 1}^n E_s,
\end{equation}
and
\begin{eqnarray}
E_s & = &
\left( \begin{array}{cc} 1 - \epsilon & \epsilon  \\     0     &    1   \end{array} \right)
\left( \begin{array}{cc} 1 - \eta_s & \eta_s  \\     \eta_s     &     1 - \eta_s   \end{array} \right)  \nonumber\\
       & = &
\left( \begin{array}{cc} 1 - \epsilon - \eta_s + 2\epsilon\eta_s  & \epsilon + \eta_s - 2\epsilon\eta_s  \\
	                                                                                                     \eta_s          &  1 - \eta_s     \end{array} \right).  \label{Mdef}
\end{eqnarray}

Eqs.~(\ref{probK}) and (\ref{probLgivenK}) imply that probability that a random read will be the pattern $\ell$ is  
\begin{equation}
\Prob(L = \ell) = \sum_{k=1}^{2^n} \theta_k M_{k \ell} = \phi_\ell, 
\end{equation}
say.
Assuming each read to be independent, for an experiment with a given total
number of reads $N_{\rm read}$ the observed set of read counts represented by
the random variable $Y_\ell$ has a multinomial distribution: 
\begin{equation}
\Prob(Y_\ell = y_\ell | \phi) =  \frac{N_{\rm read} !}{y_1! y_2! \ldots y_{2^n}!} {\phi_1}^{y_1} \ldots {\phi_{2^n}}^{y_{2^n}}.  
\end{equation}

In a recent applications note, \cite{Akman14} develop a statistical model for
the distribution of the number of reads which register as being methylated in a
pooled set of bisulphite-sequencing reads from CpG sites in a given region of a
genome.
Their model is mathematically equivalent to the $n = 1$ version of the above
model, and as such can be simplified to a single binomial distribution (see
Supplementary Material).  

\subsection*{Parameter estimation}
\label{sec:paramEst}

\begin{figure}
\centering
\includegraphics[width=0.6\textwidth]{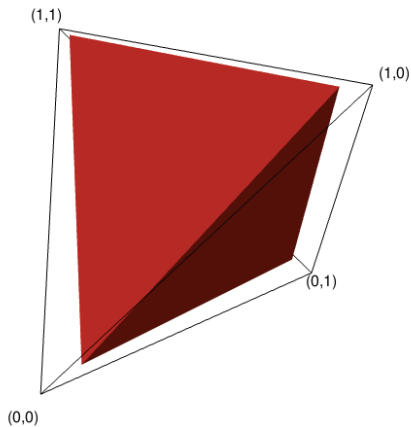}
\caption{Boundary of the allowed simplex Eq.~(\ref{thetaSimplex}) for the parameters $\theta$, (black wire frame) and corresponding 
shrunken simplex containing allowed values of $\phi$ (red tetrahedron) for the case $n = 2$ CpG sites, $\epsilon = 0.05$, $\eta_s = 0.02$.  
Numbers at the corners are the $2^n = 4$ possible methylation patterns)} 
\label{fig:tetrahedron_eps05_eta02}
\end{figure}

The parameters of the distribution over methylation profiles, $\theta_k$, are
estimated by maximising the log likelihood: 
\begin{equation}
L(\theta | y) = \sum_{k = 1}^{2^n} y_k \log\left( \sum_{j=1}^{2^n} \theta_j M_{jk} \right), 
\end{equation}
subject to the constraint that $(\theta_1, \ldots, \theta_{2^n})$ lies in the $(2^n - 1)$-dimensional simplex
\begin{equation}
S = \left\{ \theta : \sum_{k = 1}^{2^n} \theta_k = 1, \theta_k \ge 0 \right\}.       \label{thetaSimplex}
\end{equation}
One may be tempted to use the usual formula, $\hat \phi_\ell =  y_\ell/N_{\rm
read}$, for the maximum likelihood estimate of multinomial parameters, and
simply invert the matrix $M$ to recover $\hat \theta_k$.
However this will not work for any realistic data because the matrix $M$ shrinks the 
simplex to a smaller volume.  
In practice many of the $y_\ell$ are zero, which leads to a naive estimate
$\hat \phi_\ell$ on the boundary of the unshrunken simplex in $\phi$-space, and
this boundary is not included in the shrunken simplex (see
Fig.~\ref{fig:tetrahedron_eps05_eta02} for the $n = 2$ case, in which the
simplex is  tetrahedron).  

Instead we maximise the log likelihood over the allowed domain numerically by
using the R function {\tt constrOptim()}.
Unfortunately the performance of this function becomes prohibitively slow for
$n \gtrsim 8$ as the dimensionality of the parameter space grows exponentially.
However, we have noticed in numerical simulations that if the observed counts
$y_k$ are zero for a subset of the possible patterns, the corresponding
estimates $\hat\theta_k$ are zero (or in rare cases, very close to zero) for
the same subset.
Thus we have implemented an algorithm which only searches over that part of the
boundary of $S$ constrained by $\theta_k = 0$ for those $k$ such that $y_k =
0$.
The algorithm remains reasonably efficient on a standard desktop computer
provided the number of observed methylation patterns does not exceed about 200.
This allows analysis of most realistic datasets while still addressing the
biologically relevant question of identifying spurious methylation patterns
which are the result of incomplete methylation.  

Finally, to adjust for the fact that the function {\tt constrOptim()} is of
finite accuracy in locating the maximum of the log-likelihood, if the located
maximum $\hat\theta$ is close to the boundary of the simplex, the value of the
log-likelihood is also calculated at several nearby points on the boundary.
If this results in a log-likelihood bigger than or equal to the maximum
reported by {\tt constrOptim()}, the appropriate point on the boundary is taken
as the maximum likelihood estimate, and those patterns $m_k$ for which
$\hat\theta_k \equiv 0$ are reported as being spurious reads.

\section*{Conclusions}

We have developed an algorithm for estimating the true distribution of
methylation patterns in at a genomic locus containing $n$ CpG sites.
The algorithm, based on a constrained multinomial model, accounts for
statistical variation due to incomplete bisulphite conversion and sequencing
errors.  
The analysis differs from previous treatments in that the estimated
distribution is a joint probability distribution over patterns which preserves
maximal information pertaining to interaction between different CpG sites, as
opposed to a pointwise measure of methylation at each site.  
A pointwise methylation estimate can, of course, be recovered from our
estimated distribution as a marginal distribution.  
The algorithm is implemented as the R Bioconductor package {\tt MPFE}. 

Numerical experiments with realistic synthetic data indicate that the algorithm
is able to identify the majority of the spurious observed methylation patterns,
that is, patterns which are not present in the original library but are
observed in the reads because of incomplete bisulphite conversion or sequencing
errors.
In general, our estimates are closer to the true distribution than the naive
estimates given by the relative proportion of observed read counts for almost
all patterns in each simulation (see Figures~\ref{swco}, \ref{syntheticBee},
and Table~\ref{syntheticTable}).

Application of the algorithm to biological data consisting of bisulphite
treated amplicon reads for a honeybee genomic sequence predicts that a
correspondingly high proportion of observed methylation patterns in real data
may indeed be spurious.
However, our results also reveal an important number of real methylation
patterns in this biological sample.
This complexity of the methylation landscape is virtually undetectable when one
only considers position-wise methylation level, but becomes apparent through
our method.

\section*{Competing interests}
  The authors declare that they have no competing interests.

\section*{Author's contributions}
SF, CJB conceived and designed the project. 
SF prepared the datasets. 
PL, CJB developed the mathematical and statistical model, 
carried out the analysis and developed the software. 
PL, SF, CJB wrote the manuscript. 
All authors read and approved the final manuscript.

\section*{Acknowledgements}

This project is supported by Australian
Research Council Discovery Grants DP120101422 and DE130101450 and National
Health and Medical Research Council Grant NHMRC525453.  We thank Ryszard
Maleszka for sharing sequencing data. \\

% if your bibliography is in bibtex format, use those commands:
\bibliographystyle{bmc-mathphys} % Style BST file
\bibliography{bmc_article}      % Bibliography file (usually '*.bib' )

% or include bibliography directly:
% \begin{thebibliography}
% \bibitem{b1}
% \end{thebibliography}

\section*{Additional Files}
  \subsection*{Additional file 1 --- Comparison with Akman et al. (2014)}
  Analysis of the relationship of the statistical model used by Akman et al. \cite{Akman14}
  to our statistical model.

\end{document}